\documentclass{sig-alternate}
\begin{document}

%
\conferenceinfo{JFO 2009}{December 3-4, 2009, Poitiers, France}
\CopyrightYear{2009} 
\crdata{978-1-60558-842-1}  

\title{Multi-repr\'{e}sentation d'une ontologie : OWL, bases de donn\'{e}es, 
syst\`{e}mes de types et d'objets.}
%
%

\numberofauthors{2}
%

\author{
%
\alignauthor Mireille Arnoux \\
       \affaddr{Universit\'{e} de Bretagne Occidentale}\\
       \affaddr{D\'{e}partement d'informatique}\\
       \affaddr{20, Avenue Le Gorgeu 29238 Brest Cedex 3}\\
       \email{Mireille.Arnoux@univ-brest.fr}
\alignauthor Thierry Despeyroux \\
       \affaddr{INRIA}\\
       \affaddr{Centre de recherche Paris-Rocquencourt}\\
       \affaddr{B.P. 105 - 78153 Le Chesnay Cedex, France }\\ 
       \email{Thierry.Despeyroux@inria.fr}
}

\date{18 June 2009}
\maketitle
\begin{abstract}
Due to the emergence of the semantic Web and the increasing need to
formalize human knowledge, ontologie engineering is now an important
activity. But is this activity very different from other ones like
software engineering, for example ?

In this paper, we investigate analogies between ontologies on one
hand, types, objects and data bases on the other one, taking into
account the notion of evolution of an ontology. We represent a unique
ontology using different paradigms, and observe that the distance
between these different concepts is small. We deduce from this
constatation that ontologies and more specifically ontology
description languages can take advantage of beeing fertilizated with
some other computer science domains and inherit important
characteristics as modularity, for example.

\end{abstract}


 {Languages}




\section{Introduction}
Les ontologies sont une mani\`{e}re de repr\'{e}senter de façon formelle la
connaissance. Un des buts principaux des ontologies est d'être partag\'{e}es
entre un groupe de personnes pour fixer une terminologie et les
relations entre concepts, aussi bien pour une utilisation humaine que
pour une machine \cite{Gruber}.  Ces ontologies sont utilis\'{e}es par des annotations s\'{e}mantiques qui sont la base du Web semantique \cite{berners-lee:2001,Lee98}.

De plus en plus d'ontologies sont maintenant publi\'{e}es directement
sur le Web.  D'abord ``l\'{e}g\`{e}res'' (comrtant les termes d'un
vocabulaire et des relations entre ces termes) elles laissent
maintenant la place à des ontologies plus lourdes qui comportent aussi
des r\`{e}gles et des possibilit\'{e}s de raisonnement logique, et à
l'élaboration de méta-modèles (Ontology definition metamodel ODM) en
rapprochant les visions logiques (DL), structurelle (OWL) et génie
logiciel (ULM).

Cette complexit\'{e} du
d\'{e}veloppement des ontologies a amen\'{e} \`{a} parler de g\'{e}nie
ontologique comme on parle de g\'{e}nie logiciel.
Ces deux diciplines ont pour l'instant peu d'interactions, et pourtant
elles ont de nombreux points communs. De même que les objets manipul\'{e}s
par un programmes peuvent devoir (pour des raisons de performances par
exemple) s'incarner dans des structures de donn\'{e}es diff\'{e}rentes, une
ontologie, qui est le plus souvent d\'{e}crite en utilisant OWL, peut
devoir être repr\'{e}sent\'{e}e de multiples façons suivant l'utilisation
qu'on veut en faire : syst\`{e}mes de types, bases de donn\'{e}es, langages
objets. Les possibilit\'{e}s d'expres\-sion de ces diff\'{e}rents paradigmes
\'{e}tant diff\'{e}rentes, les confronter devrait permettre d'importer dans le
monde des ontologies de nouveaux concepts. Un autre point commun entre
ontologies et logiciels est la notion d'\'{e}volution et de
cons\'{e}quences d'une \'{e}volution.

Nous introduisons d'abord OWL et un extrait d'ontologie représenté dans ce langage, puis un état de l'art  sur les transformations  possibles dans les divers paradigmes évoqués. Nous proposons alors une traduction de notre exemple dans chacun d'eux. Nous soulignons les spécificités apportées et
nous allons  plus loin que la simple traduction d'un exemple statique en examinant  des cas d'évolution de l'ontologie ; au niveau base de donn\'{e}es
ou syst\`{e}mes de types et d'objets, nous montrons que l'on peut rep\'{e}rer les instances qui
deviennent incoh\'{e}rentes avec la nouvelle structure, et que,  grâce à la trace du
changement de l'ontologie OWL, nous pouvons assurer la robustesse de notre multi-représentation en  faisant évoluer les traductions.  Nous terminons en résumant les apports et limites de chaque représentation et en envisageant une synthèse des points les plus positifs.   

\section{OWL et un extrait d'ontologie}
Le langage OWL  (Web Ontology Language) est le langage de marquage s\'{e}mantique standard du web. OWL \'{e}tend RDF et RDFS  pour mieux d\'{e}crire les propri\'{e}t\'{e}s et les classes (classes 
disjointes et connecteurs logiques entre elles, caract\'{e}ristiques globales des propri\'{e}t\'{e}s comme la r\'{e}flexivit\'{e}{\ldots}).
Il admet, \`{a} c\^{o}t\'{e} de sa syntaxe abstraite, une syntaxe 
concr\`{e}te en RDF/XML (que l'on peut abréger en triplets N3 comme nous le ferons dnas la suite). 

OWL permet de raisonner et d'\'{e}tablir des inf\'{e}rences.
Le contenu principal d'une ontologie OWL  tient 
dans ses axiomes qui fournissent des informations concernant les classes et 
les propri\'{e}t\'{e}s et dans ses faits qui fournissent des informations 
concernant les individus.

Notre illustration est un extrait d'ontologie inspir\'{e} de \cite{luong:phd}. La 
figure 1 pr\'{e}sente les classes et la figure 2 
les propri\'{e}t\'{e}s sous Prot\'{e}g\'{e}/OWL \cite{protegeowl:CO-ODE}.

En suivant une m\'{e}thodologie classique pour construire  une ontologie, on place d'abord des classes primitives  en hi\'{e}rarchie simple (ronds gris de la figure 1). Voici un extrait en OWL ou N3 :
\begin{verbatim}
:Manager  rdfs:subClassOf :Person .
\end{verbatim}

\begin{figure}
\centering
\epsfig{file=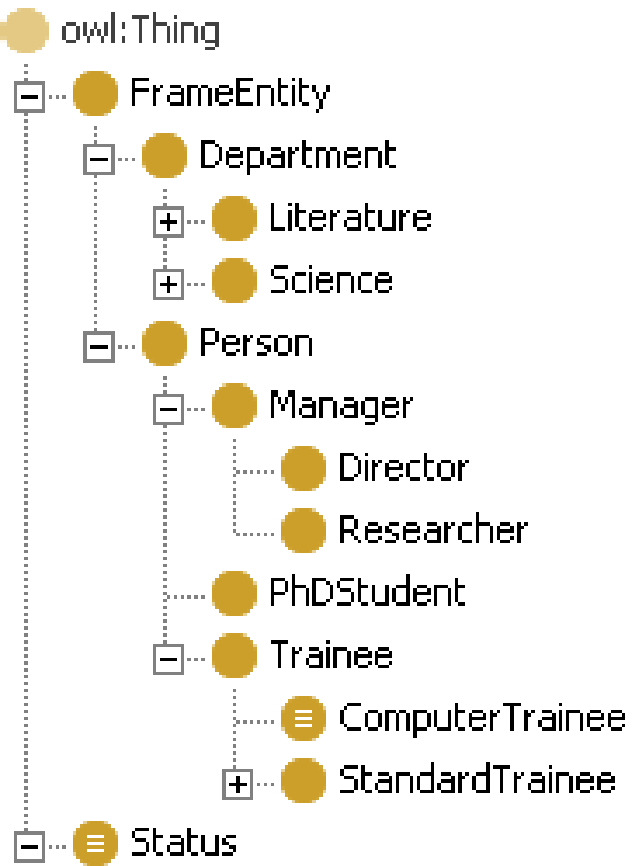, height=2.58in, width=1.98in}
\caption{Classes}
\end{figure}

Les propri\'{e}t\'{e}s sont ind\'{e}pendantes des classes  et hi\'{e}rachi\-s\'{e}es (figure 2). Leur cible est un type standard ou une classe (propri\'{e}t\'{e} objet). Les propri\'{e}t\'{e}s objet sont suppos\'{e}es multi-valu\'{e}es avec, éventuellement, un domaine et un co-domaine cible (ou range) comme dans :
\begin{verbatim}
:studyAmong a rdf:Property.

:studyAmong rdfs:domain :Trainee; 
            rdfs:range :Department.
\end{verbatim}

Une originalit\'{e} de la d\'{e}marche ontologique qui relie les instances \`{a} des classes a posteriori et permet de donner une propri\'{e}t\'{e} quelconque \`{a} un individu (à charge pour un raisonneur d'en \'{e}valuer les cons\'{e}quences) est \`{a} prendre en compte. Dans notre exemple, choisir {\tt Trainee} pour domaine de la propriété {\tt studyAmong} am\`{e}nera \`{a} proposer tout individu ayant la propri\'{e}t\'{e} {\tt studyAmong}   comme une instance de {\tt Trainee}.

Les aspects globaux d'une propri\'{e}t\'{e} (fonctionnalit\'{e}, transitivit\'{e}) s'expriment aussi grâce à OWL ; ici {\tt work} et {\tt manage} sont fonctionnelles.
\begin{figure}
\centering
\epsfig{file=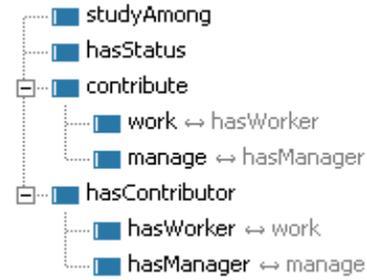, height=1.48in, width=1.98in}
\caption{Propri\'{e}t\'{e}s}
\end{figure}

On enrichit ensuite la description pr\'{e}c\'{e}dente en introduisant des classes d\'{e}finies, notamment les restrictions. Une restriction est une classe anonyme demandant que ses instances satisfassent \`{a} une restriction donn\'{e}e (cardinalit\'{e}, existence d'une valeur cible de classe donn\'{e}e) sur une propri\'{e}t\'{e}. 
Par exemple on peut imposer que {\tt studyAmong} ait au moins une valeur cible de classe {\tt Computer} :
\begin{verbatim}
[ a owl:Restriction; owl:onProperty :studyAmong
                   ; owl:someValuesFrom :Computer].
\end{verbatim}

On peut alors pr\'{e}ciser la classe {\tt ComputerTrainee} sous-classe de {\tt Trainee} en lui imposant d'être sous-classe de la classe anonyme pr\'{e}c\'{e}dente.
En fait, c'est même  une classe \'{e}quivalente \`{a} l'intersection des sur-classes (classe d\'{e}finie, dite compl\`{e}te en OWL). 

Le Web s\'{e}mantique d\'{e}centralise les instances (faits, ABoxes) conformes au sch\'{e}ma (axiomes, TBoxes) d'une ontologie. Ce sont des sources de donn\'{e}es RDF  ou des annotations d'une page Web. Dans notre exemple :
\begin{verbatim}
:r1 a :Person.      
:r2 a :PhdStudent.
:r3 a :Manager.     
:r4 a :Manager.
:r5 a :Researcher.  
:r6  a :Researcher.
:r7 a :Director.    
:r8 a :Director.
:v1 a  :Department. 
... 
:v8 a  :Department
:r1 :work :v1.   
:r2 :work :v2. 
:r3 :work :v3.  
:r4 :manage 
:v4. :r5 
:work :v5. 
:r6 :manage 
:v6. :r7 
:work :v7. 
:r8 :manage :v8.
\end{verbatim}

\section{Évolution des ontologies}

Tout comme un programme, une ontologie est une entité vivante. Il y a
bien sur la phase de développement pendant laquelle l'ontologie est
régulièrement augmentée par de nouveaux concepts, mais il faut ensuite
suivre l'évolution non seulement des connaisances qui doivent être
formalisées mais aussi prendre en compte les contraintes
diverses liées à leur utilisation. Tout comme un programme, les
connaissances et les ontologies qui les représentent doivent être
maintenues, avec bien sur un fort impact sur tout ce qui utilise ou
dépend de ces ontologies.

L'\'{e}volution d'une ontologie est donc une situation fr\'{e}quente et 
d\'{e}licate qui demande \`{a} \^{e}tre tr\`{e}s bien d\'{e}finie \cite{flouris:change} et ses 
cons\'{e}quences sur les instances existantes doivent \^{e}tre prises en 
compte.

Nous envisageons ici à titre d'exemple deux cas simples (successifs)
d'\'{e}volution de l'ontologie, dont nous suivrons les conséquences
dans les différents modes de représentation.

\emph{Modification 1:  changement de domaine}

{\tt manage} n'a plus pour domaine {\tt Manager} mais {\tt Director}.   

\emph{Modification 2:  suppression de classe}

le sous-type {\tt Manager} de {\tt Person} est supprim\'{e}.

\section{Traduction d'ontologies dans\\divers langages : \'{e}tat de l'art}

Le niveau le plus g\'{e}n\'{e}ral de dualit\'{e} de repr\'{e}sentation d'une ontologie est entre son expression en g\'{e}nie ontologique et en g\'{e}nie logiciel (pratiquement entre l'usage d'un langage d'ontologies comme OWL de syntaxe RDF  et un langage de mod\`{e}les comme UML dans le cadre d'Eclipse). Dans \cite{hillairet:EMFRDF}    est pr\'{e}sent\'{e}e notamment une 
correspondance entre une ontologie et un mod\`{e}le objet. Elle se rapproche de la correspondance ORM entre 
mod\`{e}le objet et base de donn\'{e}es relationnelle d\'{e}j\`{a} 
utilisable avec Eclipse.

Pour ce qui est du passage dans un langage objet, il est souvent motiv\'{e} par l'acc\`{e}s aux  instances mais traduit aussi l'ontologie.
ActiveRDF \cite{oren:activeRDF} permet d'utiliser des donn\'{e}es 
s\'{e}mantiques dans les langages orient\'{e}s objet. L'accent est surtout 
mis sur les adaptateurs g\'{e}n\'{e}riques SPARQL ou sp\'{e}cialis\'{e}s 
mais les diff\'{e}rences \`{a} surmonter au niveau de l'h\'{e}rita\-ge et de la consistance g\'{e}n\'{e}rale sont aussi \'{e}voqu\'{e}es. Bartalos
\cite{bartalos:object-onto} propose la cr\'{e}ation automatique de classes au sens 
orient\'{e} objet pour contenir des objets correspondant aux instances 
ontologiques. Il souligne que les classes n'ont pas la pr\'{e}tention de 
couvrir toute la complexit\'{e} de l'ontologie ni de 
servir pour un quelconque raisonnement. 
OntoJava \cite{eberhart:javasqlinference} traduit des ontologies \'{e}crites avec 
Prot\'{e}g\'{e} et des r\`{e}gles en RuleML dans une base d'objets avec un 
moteur de r\`{e}gles.
Kalyanpur \cite{kalyanpur:owl-java} note d'abord la richesse s\'{e}mantique de OWL 
et cerne les diff\'{e}rences entre la logique de description et les langages 
orient\'{e}s objet pour traduire le mieux possible le mod\`{e}le ontologique 
en Java ; l'environnement de d\'{e}veloppement Java a l'avantage de tracer et 
montrer les erreurs sur l'application ou l'ontologie. De plus une 
documentation est automatiquement cr\'{e}\'{e}e. 

Quant aux travaux \'{e}tablissant un \'{e}change entre 
ontologies et base de donn\'{e}es, ils sont tr\`{e}s divers et les correspondances propos\'{e}es sont  orient\'{e}es 
soit de la base de donn\'{e}es vers le mod\`{e}le ontologique lorsqu'il s'agit de 
rendre accessible aux internautes le web profond soit de l'ontologie vers 
une base de donn\'{e}es lorsqu'il s'agit de r\'{e}aliser un stockage 
optimis\'{e}.

Pour traduire une base de donn\'{e}es en une ontologie on utilise surtout son 
sch\'{e}ma logique ou m\^{e}me son sch\'{e}ma conceptuel \cite{cullot:dbowl}, \cite{astrova:sql-onto}.
Des langages sp\'{e}cialis\'{e}s comme D2RQ \cite{bizer:D2RQ} 
expriment les correspondances g\'{e}n\'{e}r\'{e}es automatiquement ou 
explicit\'{e}es par l'utilisateur.
L'ontologie n'est en g\'{e}n\'{e}ral pas explicitement peupl\'{e}e comme 
dans Virtuoso \cite{erling:virtuoso} o\`{u} les requ\^{e}tes du type SPARQL sont 
traduites en SQL.
Bordiga \cite{borgida:kr-db} montre la compl\'{e}mentarit\'{e} d'une ontologie vue en r\'{e}seau s\'{e}mantique et en base de donn\'{e}es : cette dernière assure la 
gestion de la concurrence, la pr\'{e}vention et la r\'{e}cup\'{e}ration des 
erreurs.
Une ontologie de domaine pr\'{e}existante peut aussi \^{e}tre coupl\'{e}e 
\`{a} une base de donn\'{e}es \cite{hu:rdbschema-onto}. Dans cette optique, o\`{u} la correspondance n'est pas une traduction syst\'{e}matique, un langage spécifique doit l'exprimer \cite{barrasa:R2O}.
Enfin,  \cite{astrova:owl-storesql} \'{e}nonce des r\`{e}gles de transformation d'une ontologie vers une base de donn\'{e}es : les tables ne se 
bornent pas \`{a} des triplets comme celles des 
entrep\^{o}ts RDF mais refl\`{e}tent au mieux les concepts et 
propri\'{e}t\'{e}s.

\begin{table*}
\begin{center}
\begin{tabular}{|p{65pt}|p{70pt}|p{65pt}|p{85pt}|p{110pt}|}
\hline
NOTION \'{E}TUDI\'{E}E & 
OWL & 
SYST\`{E}ME DE TYPES& 
LANGAGE OBJET& 
BASE DE DONN\'{E}ES \\
\hline
Taxonomie& 
Classes& 
Types & 
Classes& 
Tables \\
\hline
Hi\'{e}rarchisation& 
Subsomption& 
Sous-type& 
Sous-classe& 
Tables de m\^{e}me cl\'{e} ou encha\^{\i}n\'{e}es \\
\hline
Description \par structurelle& 
Propri\'{e}t\'{e}s& 
Op\'{e}rations& 
Attributs \par M\'{e}thodes& 
Colonnes ou \par tables de liaison \\
\hline
Description \par D\'{e}taill\'{e}e  & 
Domaine, Cible \par Restrictions sur propriétés& 
Signatures \par d\'{e}finitions& 
Corps des m\'{e}thodes & 
Contraintes(domaine, intégrité, référence)\par Triggers  \\
\hline
Lien  taxonomie - description& 
\textbf{Propri\'{e}t\'{e}s d\'{e}finies \`{a} part}& 
Op\'{e}rations li\'{e}es au type& 
M\'{e}thode  définies dans les classe\par Interface& 
Description int\'{e}gr\'{e}e au sch\'{e}ma des tables \\
\hline
Polymorphisme& 
union, Restriction de propri\'{e}t\'{e}s & 
Type \par discriminant& 
Red\'{e}finition et surcharge& 
Contraintes sur une vue ou sous-table \\
\hline
 Mod\`{e}le/ \par Instance& 
Classes/ \par Individus& 
Types/  Variables de type & 
Classes/ \par Objets& 
Structure de table/ Lignes de table (instance de relation) \\
\hline
Equivalence des concepts& 
\textbf{Same} \par \textbf{ConceptAs}& 
Seulement types compatibles& 
Nom de classe unique& 
Eventuellement vue = table \\
\hline
Identit\'{e} des individus& 
\textbf{SameAs mais IRI unique}& 
Individus locaux& 
Objets  locaux, persistance par ORM& 
Cl\'{e} primaire ou OID en relationnel-objet \\
\hline
\end{tabular}
\label{tab1}
\caption{
Notions fondamentales
}
\end{center}
\end{table*}

\section{Vision en  système de types}

Les ontologies et les annotations sémantiques peuvent être mises en
parallèle avec les programmes qui sont aussi des systèmes
formels. Avec cette vision, les concepts deviennent des types et la
subsomption  une inclusion de types.  Dans ce contexte, les
propriétés se rapprochent des fonctions et donc des signatures qui
définissent leurs domaines et co-domaines.

Notre ontologie exemple peut maintenant se définir par :

\begin{verbatim}
Person, PhdStudent, Trainee, ComputerTrainee, 
  Manager, Researcher, Director, 
  Department... : type;
PhdStudent <= Person;
Trainee <= Person; 
ComputerTrainee <= Trainee
Manager <= Person; 
Researcher <= Manager; 
Director <= Manager;
work : Person ->  Department;
manage : Manager -> Department;
\end{verbatim}

Le signe {\tt <=} denote l'inclusion de type.

Les annotation sémantiques peuvent maintenant être vues comme des
expressions dans un langage de programmation. Les instances deviennent
des objets (disons des constantes) dont le type peut être au choix
inféré ou déclaré. Sachant que les langages de programmations à typage
fort permettent un meilleur contrôle des types, en générant autant de
messages d'erreur que possible, nous préfèrerons déclarer le type des
objets.

\begin{verbatim}
r1 : Person; 
r2 : PhdStudent; 
r3, r4 : Manager;
r5, r6 : Researcher; 
r7, r8 : Director;
v1, ..., v8  : Department;
\end{verbatim}

\emph{Modification 1 : {\tt manage} a pour domaine {\tt Director}}. 
  
Les objets\ 
{\tt r4} et {\tt r6}  sont utilisés dans les annotations

{\tt manage(r4, v4)} et 
{\tt manage(r6, v6)}.

Or {\tt r4} et {\tt r6}
ne sont pas de type {\tt Director} et sont donc non conformes à la signature de {\tt manage}.\\

\emph{Modification 2 : le sous-type {\tt Manager} de {\tt Person} est supprim\'{e}}.

En applicant les modifications proposées à
notre ontologie exemple, nous obtenons un nouveau système de types:

\begin{verbatim}
Person, PhdStudent, Trainee, ComputerTrainee, 
  Researcher, Director, Department : type;
PhdStudent <= Person;
Trainee <= Person; 
ComputerTrainee <= Trainee
Researcher <= Person; 
Director <= Person;
\end{verbatim}

En appliquant de façon classique un vérificateur de type à notre ensemble d'annotations, nous pouvons générer des messages d'erreur : dans la déclaration 

{\tt   r3, r4 : Manager;}  

le type (concept) {\tt Manager} n'est pas déclaré et cette déclaration n'est donc pas légale.
Les objets {\tt r3} et {\tt r4}  sont de plus  utilisés dans :

{\tt work(r3, v3)} et {\tt manage(r4, v4)}.

L'erreur de déclaration provoque donc aussi une  non conformit\'{e} aux signatures des fonctions.

Dans \cite{luong:phd}, des règles de détection des inconsistances sont proposées pour un éditeur d'ontologies. En voyant une ontologie comme un syst\`{e}me de types, la consistance de l'ontologie et des annotations est testée par le traditionnel "type-checking".

\section{Vision en langage à objets}

Les classes OWL deviennent des classes (ou des interfaces en Java car c'est le seul moyen d'obtenir un multi-héritage dans ce langage).
Analogue \`{a} {\tt owl:Thing},  une classe {\tt Thing} 
peut \^{e}tre d\'{e}finie comme base dont h\'{e}ritent toutes les autres; c'est \'{e}galement \`{a} ce niveau que l'on peut traiter la 
s\'{e}rialisation des donn\'{e}es et le recueil de l'URL.
\begin{verbatim}
Public class Thing 
{protected String objectURL ...}
\end{verbatim}
{\tt owl:SubClassOf} dicte les hi\'{e}rarchies de classes  :
\begin{verbatim}
Public class FrameEntity extends Thing {...}
Public class Manager extends Person {...}
\end{verbatim}
On obtient ainsi facilement l'analogue de la hi\'{e}rarchie simple de la figure 1. Pour aller plus loin, on a choisi l'incompatibili\-t\'{e} des statuts de {\tt Manager} et {\tt Trainee} (classes ontologiques disjointes). En objet, toute instance est g\'{e}n\'{e}r\'{e}e par une classe unique et n'est donc instance que de celle-ci et de ses sur-classes. La disjonction de {\tt Manager} et {\tt Trainee} revient à leur interdire une sous classe commune. En utilisant le multih\'{e}ritage ou en Java la mod\'{e}lisation de {\tt Manager} et {\tt Trainee} par des  interfaces I1 et I2, les notions de polymorphisme, surcharge et red\'{e}finition sont alors exploitables (en Java, I1 et I2 auront une fonction de blocage de même nom mais  retournant des types diff\'{e}rents et ceci interdira de d\'{e}finir une interface \'{e}tendant à la fois I1 et I2).

Une propri\'{e}t\'{e} P va s'associer comme champ membre \`{a} la classe C 
qu'elle a pour domaine (Thing si le domaine n'est pas pr\'{e}cis\'{e}). 
Il n'y a pas de diff\'{e}rence fondamentale entre les attributs membres qui 
repr\'{e}sentent des propri\'{e}t\'{e}s type de donn\'{e}es ou objet ni 
entre les attributs mono-valu\'{e}s ou multi-valu\'{e}s~: il suffit 
d'employer le type (éventuellement polymorphe) ou la classe voulue et d'utiliser une collection 
ad\'{e}quate pour les valeurs multiples.
La notion d'accesseur (set et get) permet d'affecter une ou des valeurs 
\`{a} une propri\'{e}t\'{e} ou de lire ces valeurs. 
\begin{verbatim}
Public class Person extends FrameEntity 
{ protected Department work ...
  public void setwork (Department d) {...}
  public Department getwork () {...} }
\end{verbatim}

Une restriction de propri\'{e}t\'{e} ne peut pas faire appara\^{\i}tre une sur-classe indépendante comme en OWL car, dans le monde de la 
programmation objet, les propri\'{e}t\'{e}s sont directement attach\'{e}es 
aux classes. Un proc\'{e}d\'{e} plus 
op\'{e}rationnel doit \^{e}tre utilis\'{e} : pour chaque 
propri\'{e}t\'{e} \`{a} restreindre dans une classe, un processus d'écoute (listener) surveille les accès au membre qu'est la propri\'{e}t\'{e}. Cette solution est tr\`{e}s souple et permet aussi de valider ou invalider 
la contrainte dynamiquement. Elle peut s'appliquer dans notre exemple pour 
{\tt ComputerTrainee} avec sa restriction {\tt someValuesFrom}~:
\begin{verbatim}
Public class ComputerTrainee 
{ protected List studyAmong
  // le listener Test sera enregistré sur 
  // les accesseurs à ce membre ...}

Public Class Test implements PropertyChangeListener
{ void surveillerPropertyChange
   (PropertyChangeEvent evt) 
   // si aucune valeur de evt n'est Computer c'est 
   // incorrect et une exception est levée }
\end{verbatim}

\emph{Modification 1 : {\tt manage} a pour domaine {\tt Director}}.

Les instances inconsistantes  ont une valeur dans le 
champ {\tt manage} et ne sont pas de classe {\tt Director}. 
Pour limiter ensuite l'usage du membre {\tt manage} \`{a} {\tt Director}, il suffit de 
supprimer (red\'{e}finition \`{a} erreur) l'accesseur {\tt set} de ce champ dans 
{\tt Manager} et {\tt Researcher}.

\emph{Modification 2 : le sous-type {\tt Manager} de {\tt Person} est supprim\'{e}}.

Les instances inconsistantes
r\'{e}pondent {\tt Manager} au message {\tt getClass} ou, plus simplement sont  dans la variable de classe gardant les r\'{e}f\'{e}rences aux instances cr\'{e}\'{e}es comme ci-dessous :
\begin{verbatim}
public class Manager{
 private static ArrayList tous = new ArrayList();
 public Manager () 
  {tous.add(new WeakReference(this));} ...}
\end{verbatim}
Pour emp\^{e}cher de cr\'{e}er d\'{e}sormais des instances de {\tt Manager}, sa 
m\'{e}thode de cr\'{e}ation est bloqu\'{e}e.

\section{Traduction en  base de donn\'{e}es}

\begin{table*}
\begin{center}
\begin{tabular}{|p{65pt}|p{90pt}|p{65pt}|p{85pt}|p{95pt}|}
\hline
CRIT\`{E}RE& 
OWL & 
SYST\`{E}ME DE TYPES  & 
LANGAGE OBJET& 
BASE DE DONN\'{E}ES \\
\hline
Utilisation g\'{e}n\'{e}rale& 
\textbf{Langage descriptif} \par \textbf{Editeurs sp\'{e}cialis\'{e}s} \par \textbf{Raisonneur}& 
Programmation imp\'{e}rative et typ\'{e}e& 
Programmation imp\'{e}rative, objet, \'{e}v\`{e}nementiel& 
Description et stockage, Modules \'{e}v\`{e}nementiels ou proc\'{e}duraux \\
\hline
Population du mod\`{e}le& 
\textbf{Limit\'{e}e ou s\'{e}par\'{e}e (annotation) } \par \textbf{Rattach\'{e}e aux mod\`{e}les parfois a posteriori}& 
R\'{e}duite~: variables d\'{e}clar\'{e}es ou non& 
Toute instance est cr\'{e}\'{e}e par une classe pr\'{e}cise (m\'{e}thode de cr\'{e}ation invoqu\'{e}e)& 
Essentielle~: stockage performant \\
\hline
\'{E}tendue de population& 
\textbf{Monde}  \textbf{ouvert}& 
Description ferm\'{e}e& 
Ferm\'{e}e mais propri\'{e}t\'{e}s facultatives possibles& 
Monde ferm\'{e} \\
\hline
Niveau d'implantation& 
Tout est visible~ mais les classes d\'{e}finies sont plut\^{o}t \`{a} l'usage du raisonneur& 
Interface, \par signature& 
Sp\'{e}cification et corps& 
Optimisations, \par Vues  \\
\hline
Prise en compte des contraintes et gestion de la coh\'{e}rence& 
Raisonneur logique (consistance, taxonomie, type inf\'{e}r\'{e})& 
Contr\^{o}le de type   statique ou dynamique. \par Compilation& 
Messages invoquant les m\'{e}thodes \par Listener& 
V\'{e}rification du mod\`{e}le (normalisation) \par Contr\^{o}le \'{e}v\`{e}nementiel \\
\hline
R\`{e}gles de d\'{e}duction, \par Reclassement& 
\textbf{Raisonneur et moteur logique d'inf\'{e}rence}& 
Programmes explicites& 
Programmes explicites& 
Vues, requ\^{e}tes (bases d\'{e}ductives) \\
\hline
D\'{e}tection d'inconsistances& 
\textbf{Raisonneur}& 
Compilateur& 
Compilateur,  Interpr\'{e}teur,  exceptions& 
SGBD, \par int\'{e}grit\'{e} \\
\hline
Int\'{e}gration R\'{e}utilisation& 
Import et Merging& 
Modules& 
Importation& 
Vues  externes,  ETL, \par Entrep\^{o}t de donn\'{e}es \\
\hline
\end{tabular}
\label{tab2}
\caption{
Mise en {\oe}uvre
}
\end{center}
\end{table*}

Les classes OWL correspondent g\'{e}n\'{e}ralement \`{a} la d\'{e}finition 
de tables entit\'{e}s. 

Les propri\'{e}t\'{e}s OWL mono-valu\'{e}es sont alors les champs de ces tables. Un cas 
particulier est l'identificateur cl\'{e} primaire qui n'est cependant pas une r\'{e}f\'{e}rence absolue comparable \`{a} une URI.
Les propri\'{e}t\'{e}s  mono-valu\'{e}es vers un objet sont des champs avec une contrainte r\'{e}f\'{e}rence. 
\begin{verbatim}
CREATE TABLE Person
(IDPerson INTEGER PRIMARY KEY,
work INTEGER REFERENCES Department);
\end{verbatim}

Les autres propri\'{e}t\'{e}s (comme {\tt StudyAmong}) donnent  des tables associations :
\begin{verbatim}
CREATE TABLE StudyAmong
(IDTrainee INTEGER REFERENCES Trainee,
IDDepartment INTEGER REFERENCES Department,
PRIMARY KEY (IDTrainee, IDDepartment))
\end{verbatim}

La traduction de sous-classes en tables s\'{e}par\'{e}es peut se faire par plusieurs m\'{e}thodes. Table et sous-table peuvent avoir même cl\'{e} ou s'encha\^{\i}ner par des colonnes r\'{e}f\'{e}rentielles, comme ici la table {\tt Person} et sa sous-table {\tt Manager} :
\begin{verbatim}
Table Person(IDPerson, 
SCPerson, DISManagerTraineePhdStudent, REFwork)

Table Manager(IDManager,
 SCManager, DISResearcherDirector, REFmanage)
\end{verbatim}
Une ligne de la table {\tt Person} ne valorisant pas les colonnes  {\tt SCPerson} (référentielle) et 
{\tt DISManagerTraineePhdStudent} d\'{e}crit une 
personne sans plus de pr\'{e}cision :
\begin{verbatim}
Person(r1,,,v1)
\end{verbatim}
Une ligne de la table {\tt Person} valorisant la colonne {\tt SCPerson} (ici à {\tt r'3}) et 
{\tt DISManagerTraineePhdStudent} (à {\tt manager})  doit \^{e}tre compl\'{e}t\'{e}e par la ligne (ici de cl\'{e} {\tt r'3}) de {\tt Manager} :
\begin{verbatim}
Person(r3,r'3,manager,v3) Manager(r'3,,,)
\end{verbatim}
La colonne {\tt v3} r\'{e}f\'{e}rencie le d\'{e}partement ({\tt Department}) o\`{u} travaille la personne 
{\tt r3}. On peut envisager qu'une personne de la sous-classe {\tt Manager} ne travaille 
que dans certains d\'{e}partements et l'exprimer par une restriction en OWL. 
Cette restriction s'\'{e}crira par un 
d\'{e}clencheur sur insertion d'instance {\tt create trigger before insert} qui v\'{e}rifiera le d\'{e}partement 
r\'{e}f\'{e}renc\'{e} 

Les classes OWL  intersection, 
union ou compl\'{e}mentaire ne sont pas  d\'{e}finies comme 
tables mais plut\^{o}t comme vues. Des d\'{e}clencheurs (triggers) de type {\tt instead of} 
pourront faire une insertion dans les tables supportant les vues.
Les d\'{e}clencheurs traduisent aussi des axiomes exprimant des 
contraintes sur les extensions des concepts : 
{\tt owl:DisjointWith} par exemple 
sera assur\'{e} par un d\'{e}clencheur contraignant l'ajout ou la 
modification.
Les faits sont donc les lignes des tables :
\begin{verbatim}
Person(r1,,,v1)
Person(r2,r'2,phdstudent,v2) PhdStudent(r'2)
Person(r3,r'3,manager,v3) Manager(r'3,,,)
Person(r4,r'4,manager,) Manager(r'4,,,v4)
Person(r5,r'5,manager,v5) 
 Manager(r'5,r''5,researcher,) Researcher(r''5)
Person(r6,r'6,manager,) 
 Manager(r'6,r''6,researcher,v6) Researcher(r''6)
Person(r7,r'7,manager,v7) 
 Manager(r'7,r''7,director,) Director(r''7)
Person(r8,r'8,manager,) 
 Manager(r'8,r''8,director,v8) Director(r''8)
\end{verbatim}

\emph{Modification 1 : {\tt manage} a pour domaine {\tt Director}}.

Les lignes inconsistantes (r4 et r6) valorisent {\tt REFmanage} sans s'enchaîner à des lignes de {\tt Director} :
\begin{verbatim}
SELECT * from Manager where REFmanage IS NOT NULL 
       and DISResearcherDirector != director
\end{verbatim}

\emph{Modification 2 : le sous-type {\tt Manager} de {\tt Person} est supprim\'{e}}.

Les instances inconsistances ({\tt r3} et {\tt r4}) sont les lignes de {\tt Manager} avec {\tt SCManager} non valu\'{e} (non spécialisées) :
\begin{verbatim}
SELECT * FROM Manager where SCManager IS NULL
\end{verbatim}

Pour faire évoluer la base de données, il suffit d'ajouter les contraintes duales des requ\^{e}tes SQL précédentes  (un avantage de placer des contraintes ou de garder des vues est la 
possibilit\'{e} de valider ou invalider l'\'{e}volution) :
\begin{verbatim}
C1) ALTER TABLE Manager ADD CONSTRAINT
 CHECK(REFmanage IS NULL 
       or DISResearcherDirector = director)

C2) ALTER TABLE Manager ADD CONSTRAINT
 CHECK(SCManager IS NOT NULL)
\end{verbatim}

\section{Conclusion et perspectives}

 Les  tables 1 et 2 résument les possibilit\'{e}s d'expression offertes pour représenter une ontologie par les divers paradig\-mes étudiés et les comparent. OWL se distingue nettement sur les points suivants :

- OWL est un langage de description des donn\'{e}es (propri\'{e}t\'{e}s  autonomes, identit\'{e} de concepts ou d'individus) 

- OWL est utilis\'{e} pour donner une s\'{e}mantique au contenu du Web (monde ouvert)

- OWL s'associe  \`{a} un raisonneur en logique~DL.
 
Ces caractéristiques  de OWL sont indispensables pour la création initiale d'une ontologie. Cependant, dans une phase ultérieure d'utilisation, nous avons vu que sa représentation par types, objets ou tables intervient fréquemment.

-	Les types objets ou tables sont clairement déclarés et vérifiés (alors que les classes apparaissent dispersées en OWL, sans véritables déclarations), les instances leur sont attachées et les compilateurs et SGBD sont très efficaces.

-	Les agents et les applications du Web bénéficient alors des aspects de programmation évènementielle que sont les processus d'écoute (listener) en objet ou les déclencheurs (trigger) en base de données.

-	La modularité est très bien gérée dans le monde de la programmation et des types qui a fait l'objet de nombreuses recherches: importation, renommage, portée des identificateurs, polymorphisme et utilisation des types comme paramètres.

Les éditeurs d'ontologies et les outils d'annotation s'effor\-cent d'intégrer certains de ces aspects. La manipulation des concepts et des faits est facilitée ainsi que leur exportation.  Ce point peut être amélioré par une meilleure approche des diverses représentations et c'est une des perspectives de notre travail.

La taille des ontologies usuelles est le plus gros problème et la modularité est indispensable au niveau du développement comme au niveau  du déploiement. M. Jarrar \cite{jarrar:ontoengineering} propose une modularisation par sujet ou par usage au niveau conceptuel. Ici encore le monde orienté objet a son influence car les modules sont décrits en ORM (langage graphique basé sur des objets et des rôles très employé pour la modélisation des systèmes d'information). Nous voyons ici une rencontre entre le génie ontologique et le génie logiciel que nous souhaitons approfondir.
Enfin, au niveau du déploiement, une démarche usuelle \cite{schlicht:module} consiste à voir une base de connaissance comme une collection de plus petites bases disposant chacune d'un raisonneur dont les résultats sont propa\-gés pour avoir une solution globale.  Une autre option est choisie dans \cite{volz:view} où sont construites des vues sur des classes et des vues sur les propriétés avec RQL et une implantation dans le serveur KAON \cite{berlin:kaon}. Nous préconisons plutôt ces solutions  plus inspirées des autres représentations : modules en objet, vues en bases de données.  

Comme nous le voyons, les divers systèmes de représentation des
ontologies ont chacun leurs avantages, sans parler des questions
d'efficacité lors de la conception ou lors de l'utilisation.  Il ne
s'agit donc pas de remplacer l'un par un autre. Par exemple, la
représentation en système de types pourrait apporter une meilleure
compréhension de la modularité, mais ne permettrait pas dans l'état
d'aborder les contraintes de cardinalité, courantes dans les
ontologies. La solution pour augmenter l'expressivité des ontologies
serait donc d'une part de concevoir un langage de description plus riche
que les langages existants
qui prendrait en compte les résultats d'autres de domaines, d'autre
par de comprendre comment compiler ces descriptions pour en obtenir
des implémentations efficaces en fonction des utilisations projetées.

{\vskip 1.5cm}


\balancecolumns

\end{document}